# Prediction of novel stable compounds in the Mg-Si-O system under exoplanet pressures

Haiyang Niu[1,2], Artem R. Oganov[3, 1, 4, 5, *], Xing-Qiu Chen[2,**], and Dianzhong Li[2]

[1]*Moscow Institute of Physics and Technology, 9 Institutskiy Lane,*

*Dolgoprudny city, Moscow Region 141700, Russia*

[2]*Shenyang National Laboratory for Materials Science, Institute of Metal Research,*

*Chinese Academy of Sciences, Shenyang 110016, China*

[3] *Skolkovo Institute of Science and Technology, Skolkovo Innovation Center, 3 Nobel St., Moscow*

*143026, Russia*

[4] *Department of Geosciences, Center for Materials by Design, and Institute for Advanced*

*Computational Science, State University of New York, Stony Brook, NY 11794-2100.*

[5] *School of Materials Science, Northwestern Polytechnical University, Xi'an 710072, China*

(Dated: October 11, 2015)

\* *artem.oganov@stonybrook.edu*    \*\* *xingqiu.chen@imr.ac.cn*

# Abstract

The Mg-Si-O system is the major Earth and rocky planet-forming system. Here, through quantum variable-composition evolutionary structure explorations, we have discovered several unexpected stable binary and ternary compounds in the Mg-Si-O system. Besides the well-known $SiO_2$ phases, we have found two extraordinary silicon oxides, $SiO_3$ and SiO, which become stable at pressures above 0.51 TPa and 1.89 TPa, respectively. In the Mg-O system, we have found one new compound, $MgO_3$, which becomes stable at 0.89 TPa. We find that not only the $(MgO)_x$ $(SiO_2)_y$ compounds, but also two $(MgO_3)_x$ $(SiO_3)_y$ compounds, $MgSi_3O_{12}$ and $MgSiO_6$, have stability fields above 2.41 TPa and 2.95 TPa, respectively. The highly oxidized $MgSi_3O_{12}$ can form in deep mantles of mega-Earths with masses above 20 $M_\oplus$ ($M_\oplus$:Earth's mass). Furthermore, the dissociation pathways of pPv-$MgSiO_3$ are also clarified, and found to be different at low and high temperatures. The low-temperature pathway is $MgSiO_3 \Rightarrow Mg_2SiO_4 + MgSi_2O_5 \Rightarrow SiO_2 + Mg_2SiO_4 \Rightarrow MgO + SiO_2$, while the high-temperature pathway is $MgSiO_3 \Rightarrow Mg_2SiO_4 + MgSi_2O_5 \Rightarrow MgO + MgSi_2O_5 \Rightarrow MgO + SiO_2$. Present results are relevant for models of the internal structure of giant exoplanets, and for understanding the high-pressure behavior of materials.

# **Introduction**

Several astonishing discoveries have been recently achieved in planetary science, e.g, the discovery of super-Earth Gliese 830c [1]. This planet weighs at least 5 $M_⊕$ ($M_⊕$: Earth's mass) and is the nearest candidate for habitable planet so far; a new type of planet, Kepler-10c, weighing 17 times as much as Earth, is also found to be a rocky planet [2]. Such a planet was previously believed to be impossible to form, because anything so heavy would grab hydrogen gas as it grew, and become a Jupiter-like gas giant. For now, this planet is the biggest rocky planet ever discovered, much bigger than previously discovered "super-Earths" (with masses 1 to 10 $M_⊕$), making it a "mega-Earth" (with masses over 10 $M_⊕$) [2]. These breakthroughs emphasize the importance of the exploration of internal structure and mineralogy of super-Earths and mega-Earths.

After the mysterious anomalies at the Earth's D" layer have been at least partly explained by the discovery of the new mineral phase post-perovskite (pPv) $MgSiO_3$ [3, 4], one wonders whether phase transitions exist in $MgSiO_3$ under further compression, which is the key information to understand and model the internal structure of exoplanets. It was first reported that pPv-$MgSiO_3$ will decompose into $MgO$ and $SiO_2$ [5] under high pressure. However, with discovery of two new silicates, $MgSi_2O_5$ [6] and $Mg_2SiO_4$ [7], the dissociation pathway of pPv-$MgSiO_3$ became a complex three-step process at zero Kelvin: pPv-$MgSiO_3$ first decomposes into $Mg_2SiO_4$ and $MgSi_2O_5$ at 0.77 TPa, then $MgSi_2O_5$ breaks down into $Mg_2SiO_4$ and $SiO_2$ at 1.25 TPa, eventually $Mg_2SiO_4$ dissociates into $MgO$ and $SiO_2$ at 3.09 TPa. However, the effect of temperature on the stability of $Mg_2SiO_4$ and $MgSi_2O_5$, which is extremely important to evaluate the stability of minerals in exoplanet mantles, has not been considered.

Recently numerous counterintuitive compounds have been discovered under pressure. For instance, in Li-H system, besides "normal" LiH, new "counterintuitive" compounds $LiH_2$, $LiH_6$ and $LiH_8$ are predicted to be stable under pressure [8]; moreover, experimental synthesis and characterization confirm the existence of unexpected Na-Cl compounds (such as $Na_3Cl$ and $NaCl_3$) [9]; what's more, magnesium oxide (MgO), one of the most abundant phases in the Earth mantle, was long believed to be the only binary compound in the Mg-O system. Nevertheless, two extraordinary compounds, $MgO_2$ and $Mg_3O_2$ have been discovered to be stable above 116 GPa and 500 GPa, respectively [10]. These fascinating discoveries inspired us to explore possible stable binary and ternary compounds in the Mg-Si-O system.

In this work, we have performed comprehensive structure searches and investigations of the Mg-Si-O system in the pressure range 0.5-3 TPa. Due to the complexities of the ternary system, the Mg-Si, Si-O and Mg-O bounding binaries are discussed first. All of the ternary stable compounds (including the stable compounds discovered in this work) fall into the pseudo-binary $MgO$-$SiO_2$ and $MgO_3$-$SiO_3$ joins. Hence, we discuss ternary compounds in these two pseudo-binary systems separately. Lattice dynamics calculations for all the investigated structures show no imaginary vibrational frequencies, suggesting their dynamical stability throughout the pressure ranges reported here.

## Results and Discussions

Variable-composition structure searches using the USPEX code with up to 64 atoms in the unit cell at pressures ranging from 0.5 TPa to 3 TPa for the Mg-Si-O system have been carried out, identifying important low energy structures that are likely to gain stability within this chemical system. Before we talk about binary and ternary compounds in the Mg-Si-O system, crystal structures of elemental Mg, Si and O should be clarified. For elemental Mg, several phase transitions are predicted in the pressure range 0.5-3 TPa. In excellent agreement with previous studies [10, 11], our calculations demonstrate that Mg adopts the fcc structure at 0.5 TPa, then it transforms into the simple hexagonal (sh) structure at 0.76 TPa; interestingly, when pressure increases to 1.07 TPa, it transforms into the simple cubic (sc, or α-Po) structure. Elemental Si adopts the fcc structure at 0.5 TPa, in agreement with literature [12], but no phase transformation occurs in the pressure range of 0.5-3 TPa. Elemental O adopts a hexagonal *hp*8 structure at 0.5 TPa (Several similar structures are very close in enthalpy in the pressure range of 0.5 to 1.9 TPa. For more details, we refer to Ref.[13]) and then transforms into the orthorhombic *oC*16 structure at 1.9 TPa, in good agreement with literature [13].

## Mg-Si binary system

$Mg_2Si$ is the only binary compound in the Mg-Si system at ambient pressure [14]. When pressure is increased above 0.5 TPa, $Mg_2Si$ remains the only stable binary compound in the Mg-Si system, until it decomposes into Mg and Si at 1.41 TPa (see Fig. S3a in Supplementary Materials). In this pressure range, it adopts the well-known $AlB_2$-type structure (Fig. S1b).

## Si-O binary system

Even though silicon monoxide SiO can exist in the gas phase [15], no evidence shows that it can exist in the crystalline form, and the amorphous black solid form of silicon monoxide indeed is a mixture of amorphous silicon and silicon dioxide [15]. Therefore, silicon dioxide $SiO_2$ is still the only known oxide in the Si-O system. In agreement with previous work [16], pyrite-type $SiO_2$ transforms into the $Fe_2P$-type phase at 0.69 TPa. Nevertheless, if crystal structure exploration is carried out in the entire Si-O binary system, some unforeseeable structures are found. Fig. 1a demonstrates the pressure-composition phase diagram of the Si-O system. A new oxide, $SiO_3$, becomes thermodynamically stable at 0.51 TPa with the $tI$32 ($I\bar{4}$) structure. Interestingly, this $tI$32-$SiO_3$ can further transform into the $mP$16 ($P2_1/c$) structure at 0.82 TPa. As illustrated in Fig. 1b and c, both structures can be constructed by tricapped trigonal prism $SiO_9$ polyhedra, which is exactly the same coordination polyhedron as in $Fe_2P$-type $SiO_2$[16].

In order to further distinguish polyhedra in the two structures of $SiO_3$, effective coordination numbers (ECoN) [17] have been calculated. A large increase of the ECoN at the phase transition point from $tI$32 (ECoN = 7.48) to $mP$16 (ECoN = 8.05) phase can be observed in Fig. 1f, indicating that accommodation of increased coordination is the primary reason for the stability of $mP$16-$SiO_3$ compared to $tI$32-$SiO_3$. When pressure increases further, the ECoN of $mP$16-$SiO_3$ reaches 8.5, equal to the mean value of the $SiO_9$ polyhedron in $Fe_2P$-$SiO_2$[16]. Perhaps surprisingly, the Si-O distances are in the range from 1.53 to 1.95 Å in $tI$32-$SiO_2$ and 1.54 to 1.82 Å in $mP$16-$SiO_3$ at 0.7 TPa, respectively. These distances are unexpectedly long under such a high pressure, and comparable to the values (1.6 Å) in silica and silicates at ambient pressure. This phenomenon is likely a consequence of geometry, since the typical bond-length must increase in order to accommodate the dramatic increase in Si-O coordination. Therefore, the relative Si-O bond length must necessarily increase with increasing coordination as the bonding polyhedra's size expands to fill the space, a general phenomenon that is well-represented by a recently proposed coordinated hard sphere mixture model [18]. The same situation is also observed in $Fe_2P$-$SiO_2$ [16], which indicates the tendency to form highly coordinated structures instead of shrinking the Si-O distances to lower the system energy.

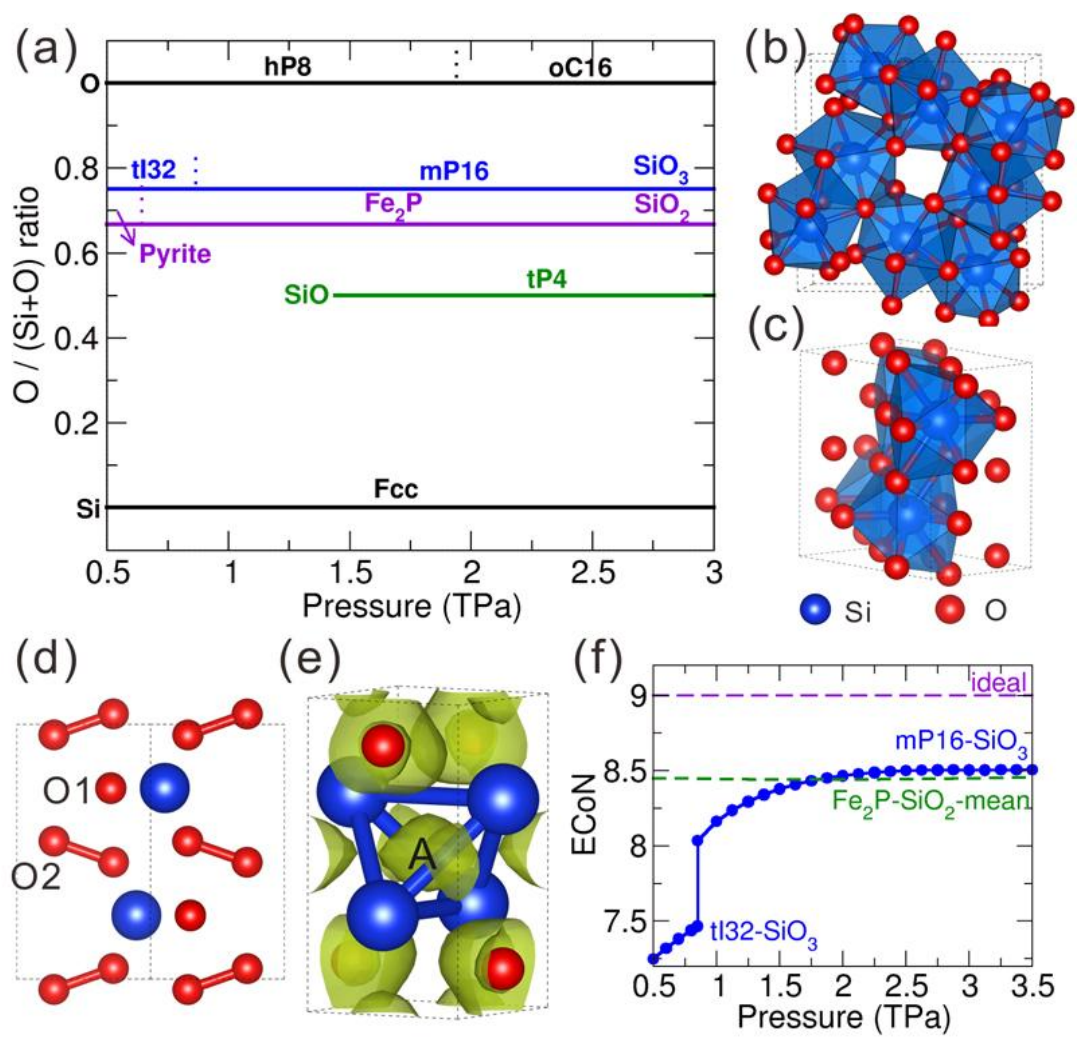

Figure 1 (a) Pressure-composition phase diagram of the Si-O system. Crystal structures of (b) $tI32$-$SiO_3$ and (c and d) $mP16$-$SiO_3$. O1 and O2 refer to two types of O atoms in $mP16$-$SiO_3$. (e) Crystal structure of $tP4$-SiO and isosurface of the electron localization function (ELF) with an isovalue of 0.65. Letter A refers to the strong interstitial electronic attractor in the $Si_4$ tetrahedron. (f) ECoN for $tI32$-$SiO_3$ and $mP16$-$SiO_3$ as a function of pressure. The mean ECoN value for $Fe_2P$-$SiO_2$ is shown by a green dashed line, and the ideal CoN of 9 is given by a purple dashed line. The densities of states of $tI32$-$SiO_3$, $mP16$-$SiO_3$, and $tP4$-SiO show that they are insulators at 0 K, see Supplementary Materials.

When pressure is raised further, stable solid silicon monoxide appears in the Si-O system with the *tP*4 structure (*P*4/*nmm*) at 1.89 TPa, see Fig. 1d and Fig. S5c. SiO crystallizes in a layered structure with Si-Si-O-O stacking order. Each Si atom is coordinated by five O atoms and eight Si atoms. Therefore, SiO retains high coordination numbers, like $SiO_2$ and $SiO_3$, despite the drop of oxygen content.

$SiO_3$ and SiO are both dynamically and thermodynamically stable, and it is still puzzling what stabilizes these exotic compounds. Based on classical chemical valence, only $SiO_2$ can be expected. To unravel the nature of these new phases, their electronic structure and chemical bonding have been analyzed.

As *tI*32-$SiO_3$ and *mP*12-$SiO_3$ display similar charge transfer and chemical bonding features, *mP*12-$SiO_3$ has been selected for the following discussion. In *mP*12-$SiO_3$ at 1 TPa, the net Bader charge [19, 20] on Si is +3.42 *e*, indicating a very large degree (~85%) of charge transfer from Si to O atoms. Based on Bader analysis, two types of O atoms exist in the *mP*12-$SiO_3$ structure (Fig.1d), the net charges on O1 and O2 are -1.63 *e* and -0.89 *e*, respectively. Therefore O1 attracts almost two electrons and attains a stable $s^2p^6$ electron configuration. Furthermore, the O-O bond distance between O2 atoms is 1.19 Å, the O-O bond distance for molecular crystal hP8-$O_2$ at 1 TPa is 1.09 Å while the non-bonding O-O distances for $MgSiO_3$ and $SiO_2$ are in the range of 1.7 Å to 2.0 Å, which clearly indicates a covalent bond and the presence of a peroxide-ion $[O-O]^{2-}$, fulfilling the octet rule. Electron Localization Function (ELF) [21] of *mP*12-$SiO_3$(Fig. S5b) confirms these conclusions: O2 atoms form peroxo-groups, while O1 atoms do not. $SiO_3$ can be classified as a "peroxide oxide", with a structural formula SiO[$O_2$], just like the recently predicted $Al_4O_7$ and $AlO_2$[22], in which $O^{2-}$ and $[O_2]^{2-}$ ions are simultaneously present.

For *tP*4-SiO at 1.5 TPa, the net charge on Si is +1.83 *e*, and the net charge on O is -1.83 *e*. Thus, O atom attains a stable electronic configuration. ELF distribution of *tP*4-SiO shows that besides accumulated electrons surrounding O atoms, we can also observe a strong interstitial electron localization in the $Si_4$ tetrahedron as marked by letter A in Fig.1e. Considering the Si-Si distance (1.86 Å) is out of the range of core-core orbital overlap, the strong interstitial electron localization is due to the formation of multicenter covalent bonds between Si atoms. Each Si atom has four nearest such electron localization regions, which accumulate two valence electrons, indeed creating an octet and explaining why each Si atom can be stabilized with two valence electrons and why SiO adopts a Si-Si-O-O ordered layered structure.

**Mg-O binary system**

Besides MgO, two novel stochiometries $MgO_2$ and $Mg_3O_2$ have recently been found to be stable under high pressure in the Mg-O system[10]. Intriguingly, if we further increase pressure, another extraordinary compound, *tP*8-$MgO_3$ with $P\bar{4}2_1m$ symmetry, becomes thermodynamically stable at 0.89 TPa as shown in the pressure-composition phase diagram of the Mg-O system (Fig. 2a). Furthermore, $Mg_3O_2$ decomposes into MgO and Mg at 0.95 TPa, while $MgO_2$ decomposes into MgO and $MgO_3$ at 1.43 TPa, and above 1.43 TPa $MgO_3$ and MgO are the only two stable magnesium oxides.

As shown in Fig. 2b, each Mg atom within MgO$_3$ has 8 nearest O neighbors (O1 atoms) forming a cubic coordination (just as in B2-MgO) and 4 second nearest O neighbors (O2 atoms). Mg and O1 atoms form a distorted fluorite-type structure, empty voids of which are stuffed with O2 atoms. According to Bader analysis, in *tP*8-MgO$_3$ at 1 TPa the net charge on Mg is +1.75 *e*, indicating the nearly complete transfer of valence electrons of Mg to O atoms. The net charges on O1 and O2 are -0.74 *e* and -0.18 *e*, respectively, while the Mg-O1 and Mg-O2 distances are 1.63 Å and 1.83 Å, respectively. Considering the O-O distance between O1 and O2 is 1.22 Å, and the O-O bond distance for molecular crystal *hP*8-O$_2$ at 1 TPa is 1.09 Å while the non-bonding O-O distances for MgSiO$_3$ and SiO$_2$ are in the range of 1.7 Å to 2.0 Å, we can conclude that two O1 atoms and one O2 atoms form a bent singly bonded [O-O-O]$^{2-}$ group. From the ELF isosurface of *tP*8-MgO$_3$ illustrated in Fig. 2c, we can also confirm the existence of [O-O-O]$^{2-}$, with a significant electronic accumulation between O1 and O2 atoms. As far as we know, this type of trioxide group is found here for the first time.

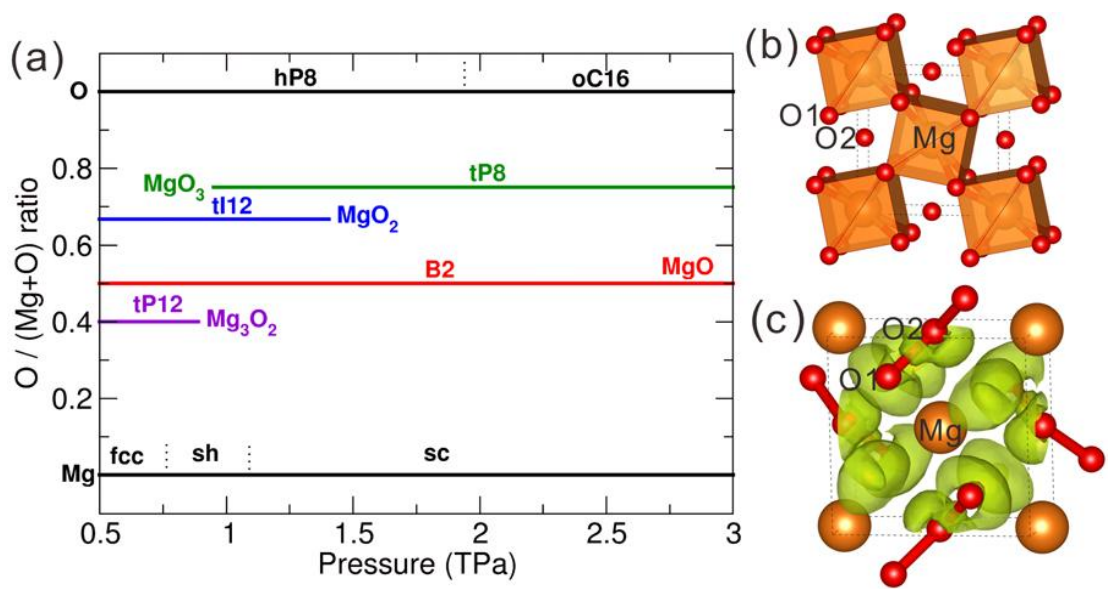

Figure 2 (a) Pressure-composition phase diagram of the Mg-O system and illustration of (b) crystal structure of *tP*8-MgO$_3$ and its (c) isosurface of the electron localization function (ELF) with an isovalue of 0.65. O1 and O2 refer to two types of O atoms in *tP*8-MgO$_3$. All Mg oxides are insulators at 0 K, see Supplementary Materials.

## Mg-Si-O ternary system

Phase diagrams of the Mg-Si-O ternary system in the pressure range 0.5-3 TPa, obtained through variable-composition crystal structure prediction for the ternary system, are shown in Fig. 3. In excellent agreement with previous works [6, 7], $Mg_2SiO_4$ and $MgSi_2O_5$ become thermodynamically stable under high pressure. We have also found two new stable ternary compounds, $MgSiO_6$ and $MgSi_3O_{12}$. The $MgO$-$SiO_2$ and $MgO_3$-$SiO_3$ pseudo-binaries contain numerous important stable compounds and are discussed in detail below.

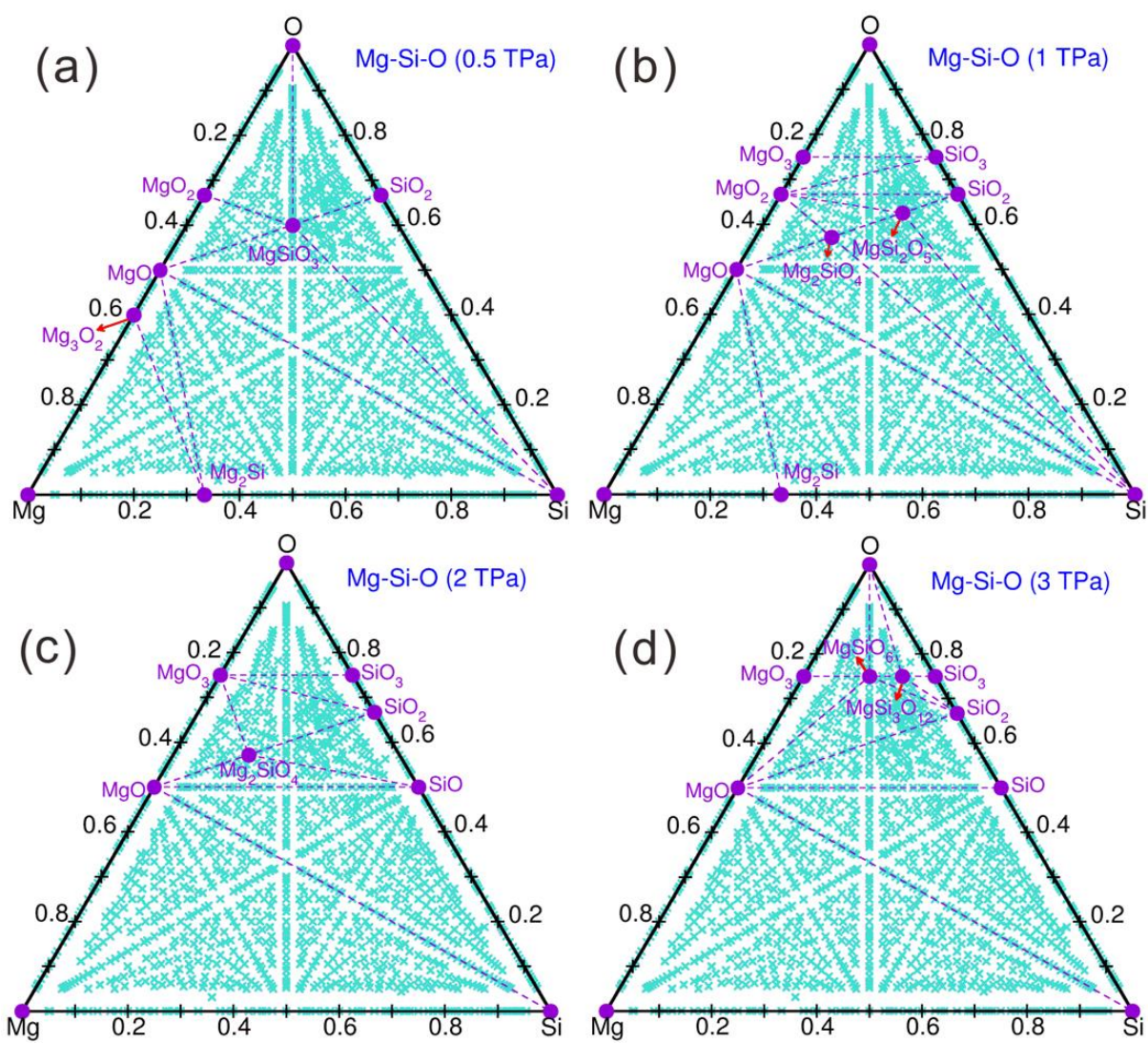

Figure 3 Mg-Si-O phase diagram at (a) 0.5 TPa, (b) 1 TPa, (c) 2 TPa and (d) 3 TPa, respectively.

**MgO-SiO$_2$ pseudo-binary system**

In good agreement with previous works [6, 7], Mg$_2$SiO$_4$ with the *tI*28 (*I*$\bar{4}$2*d*) structure and MgSi$_2$O$_5$ with the *mP*32 (*P*2$_1$/*c*) structure become thermodynamically stable at 0.51TPa and 0.63 TPa, respectively, as shown in Fig. 4a. With increasing pressure, at zero Kelvin pPv-MgSiO$_3$ decomposes into Mg$_2$SiO$_4$ and MgSi$_2$O$_5$ at 0.79 TPa, and then MgSi$_2$O$_5$ decomposes into Mg$_2$SiO$_4$ and SiO$_2$ at 1.80 TPa. Mg$_2$SiO$_4$, the last ternary compound in the MgO-SiO$_2$ pseudo-binary system, eventually decomposes into MgO and SiO$_2$ at 2.3 TPa.

Temperature, another important factor affecting stability of minerals, should be considered before developing models of the internal structure of exoplanets. Here, thermodynamic properties of these phases were investigated within the quasiharmonic approximation (QHA), using the computed phonon spectra. Previous work suggests that the *P-T* conditions of interest are within the range of validity of the QHA[5, 23]. The *P-T* phase diagram of MgSiO$_3$, as shown in Fig. 4b, is determined by comparing finite-temperature Gibbs free energies of relevant phases and phase assemblages.
In order to evaluate the electronic entropy contribution, we have calculated the electronic structures and phonon dispersions of these newly reported compounds at finite temperatures (2 kK, 5 kK, 10 kK) within the Fermi-Dirac-smearing approach [24]. We have found that all the compounds discussed in Fig.4b shows very small electronic effects at these temperatures. For instance, for the decomposition reaction of MgSiO$_3$ into Mg$_2$SiO$_4$ and MgSi$_2$O$_5$ under 0.75 TPa at 10 kK, the enthalpy changes only 0.0006 eV/atom after taking electronic entropy into consideration, and the *dP*/*dT* slope of this reaction in Fig. 4b becomes more negative, but the change is so tiny that we can safely neglect the electronic entropy contribution. Other reactions in Fig. 4b shows similar behavior.

In order to further understand this question, we have calculated the band gaps of these compounds under different pressures as listed in Table S1 in the supplementary materials. We can observe that all the compounds discussed in Fig. 3d (MgO, $SiO_2$, $MgSiO_3$, $Mg_2SiO_4$, $MgSi_2O_5$) show very big band gaps, and the electronic structures and phonon frequencies of them are not affected significantly by high temperature.

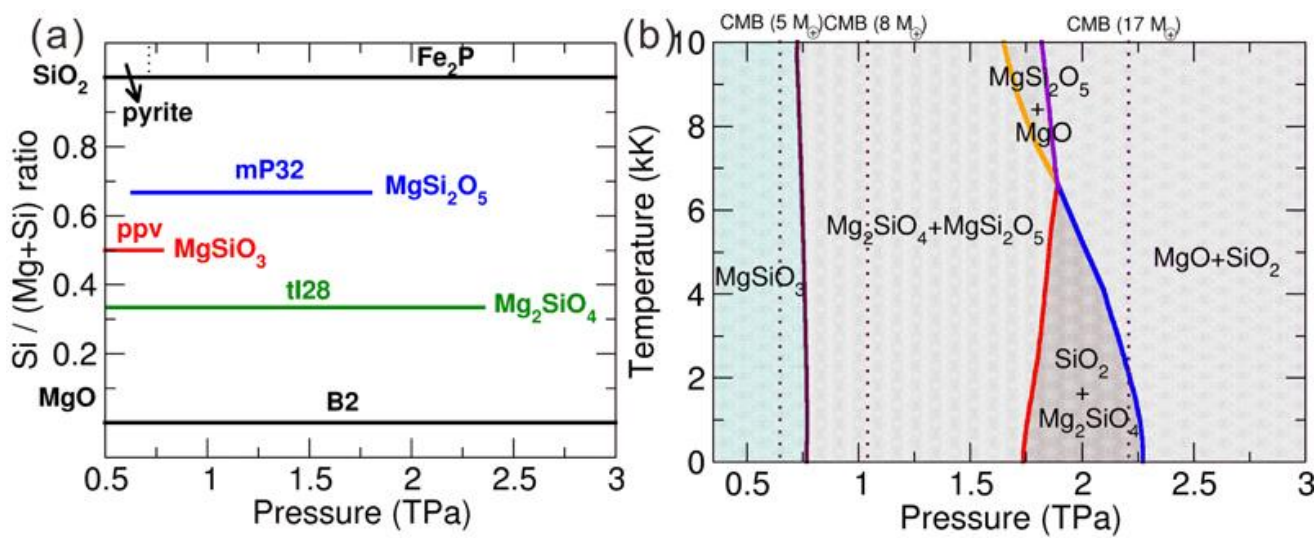

Figure 4 (a) Pressure-composition phase diagram of the pseudo-binary MgO-$SiO_2$ system. (b) P-T phase diagram of $MgSiO_3$. The core-mantle boundary (CMB) pressures of super-Earths and mega-Earths with 5, 8 and 17 $M_⊕$ are also plotted by vertical dashed lines.

As shown in Fig. 4b, the dissociation pathways of pPv-$MgSiO_3$ are different at high and low temperatures. At high temperatures (>6,610 K), $MgSiO_3$ decomposes into $Mg_2SiO_4$ and $MgSi_2O_5$, followed by decomposition of $Mg_2SiO_4$ into MgO and $MgSi_2O_5$. The last stable ternary compound in the MgO-$SiO_2$ pseudo-binary system is $MgSi_2O_5$, it eventually decomposes into MgO and $SiO_2$ at relatively high temperature well within the P-T range of mega-Earth mantles. This decomposition pathway is most likely for giant exoplanets and has not been reported before. These phase transitions and reactions are expected to impact the complex dynamics of exoplanet interiors: as exothermic transformations ($dP/dT > 0$) generally enhance heat transfer within convecting systems, while endothermic transformations ($dP/dT < 0$) decrease it[25]. As shown in Fig. 4b, the phase transition of $MgSi_2O_5$ to $Mg_2SiO_4$ and $SiO_2$ holds positive $dP/dT$ slope, and should thus help to drive convection, while all other phases transitions shown in Fig. 4b hold negative $dP/dT$ slopes, partially inhibiting convection.

**$MgO_3$-$SiO_3$ pseudo-binary system**

$MgSiO_3$, $Mg_2SiO_4$ and $MgSi_2O_5$ are traditional ordinary compounds satisfying the composition $(MgO)_x (SiO_2)_y$ (x, y: positive integers). The discovery of the novel compounds $MgO_3$, $SiO_3$ and $SiO$ suggests that other compositions may appear in the ternary system. Excitingly, we have discovered two new stable magnesium silicates which belong to the $MgO_3$-$SiO_3$ pseudo-binary system.

As shown in Fig. 5a, $MgSi_3O_{12}$ with 64 atoms in the unit cell and $cF64(Fm\bar{3})$ structure becomes stable at 2.41 TPa. By increasing pressure further, another ternary compound, $MgSiO_6$ ($cP8$, $Pm\bar{3}$) gains stability at 2.95 TPa. The two compounds share many similar structural features, as illustrated in Fig. 5b and c. Both are ordered cubic superstructures of the $Cr_3Si$-type structure. Recently[9] we have discovered a novel compound $NaCl_3$ with the $Cr_3Si$-type structure, and a related compound $NaCl_7$. This structure is stable under pressure because of high density and high coordination numbers. Mg and Si atoms in $MgSiO_6$ and $MgSi_3O_{12}$ are both icosahedrally coordinated (CN and ECoN = 12).

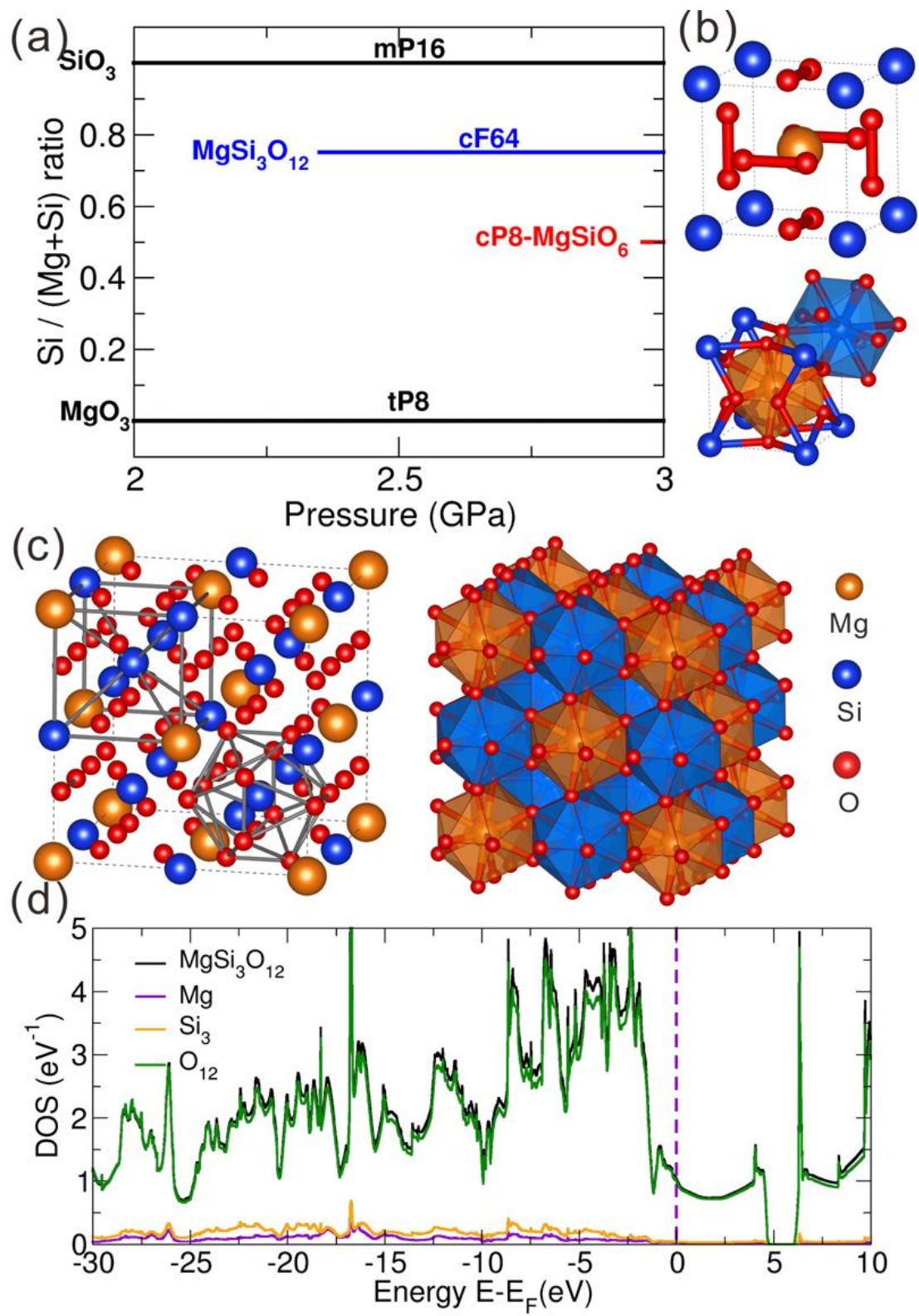

Figure 5 (a) Pressure-composition phase diagram of the pseudo-binary $MgO_3$-$SiO_3$ system and crystal structures of (b) $MgSiO_6$ and (c) $MgSi_3O_{12}$, and (d) density of states (DOS) of $cF$64-$MgSi_3O_{12}$. The density of states of $cP$8-$MgSiO_6$ is shown in Fig. S8 in Supplementary Materials.

For these new ternary magnesium silicates, we need to clarify the nature of their stability. In both compounds, one can see infinite non-intersecting O-chains along the x, y and z axes. The O-O distances in $cF$64-MgSi$_3$O$_{12}$ are in the range 1.29-1.33 Å, which are much longer than in MgSiO$_6$. Taking into account the O-O bond distance of oI16-O at 3 TPa is 1.10 Å, we can conclude that the O-O bonding in $cF$64-MgSi$_3$O$_{12}$ are much weaker than covalent singly O-O bond. From Bader analysis, for $cP$8-MgSiO$_6$ at 3 TPa, the net charge on Mg and Si are +1.59 $e$ and +3.48 $e$, respectively, while the net charge on O is -0.85 $e$, indicating the nearly complete transfer of valence electrons of Mg and Si atoms to O atoms. For $cF$64-MgSi$_3$O$_{12}$ at 3 TPa, the net charges on Mg and Si are +1.6 $e$ and +3.49 $e$, respectively, i.e. practically the same values as in $cP$8-MgSiO$_6$, while the charge on O is -1.01 $e$, which is much higher than the value (-0.85$e$) of O atom in $cP$8-MgSiO$_6$. The density of states of $cF$64-MgSi$_3$O$_{12}$ (Fig. 5d) shows that MgSi$_3$O$_{12}$ is a metal, with DOS near the Fermi level exhibiting features of a 1D-metal, which is consistent with the infinite-non-intersecting O-chains in this structure. It is worth emphasizing that all the other oxides discussed in this work are insulators, which demonstrates the unique electronic structure of $cF$64-MgSi$_3$O$_{12}$.

By adopting Fermi-Dirac-smearing approach [24], we have found that the electronic entropies of MgSiO$_6$ and MgSi$_3$O$_{12}$ are much more significant and can't be neglected. For instance, the enthalpy changes 0.10 eV/atom for MgSiO$_6$ under 3.0 TPa at 10 kK after taking electronic entropy into account. MgSiO$_6$ behaves more like a semi-conductor with band gap of 1.49 eV under 3.0 TPa, therefore bottom of the conduction band of MgSiO$_6$ becomes populated and the phonon frequencies changes at high temperature. This effect is even larger for MgSi$_3$O$_{12}$ since MgSi$_3$O$_{12}$ is a metal, the enthalpy changes 0.11 eV/atom for MgSi$_3$O$_{12}$ under 2.0 TPa at 10 kK after taking electronic entropy into account. Here we have calculated the P-T phase diagram of MgSi$_3$O$_{12}$ with and without the Fermi-Dirac-smearing. As shown in Fig. 6, the reaction from MgO3 and SiO3 to MgSi3O12 is affected significantly by electronic entropy, and the phase boundary line shifts toward lower pressure zone. For Fig. 6, we can also observe that the stability of MgSi$_3$O$_{12}$ increase with the increasing of temperature. For O-rich exoplanet, MgSi3O12 are expected to exist at high temperature and pressure. It's worth emphasizing that MgSiO6 is not stable under 3.0 TPa after considering zero-point energy, that's why MgSiO6 cannot be observed in Fig. 6. Furthermore, for metallic and semiconducting compounds predicted in this work (MgSiO$_6$, MgSi$_3$O$_{12}$), there is an intriguing possibility of their enhanced solubility in metallic iron-rich cores of exoplanets.

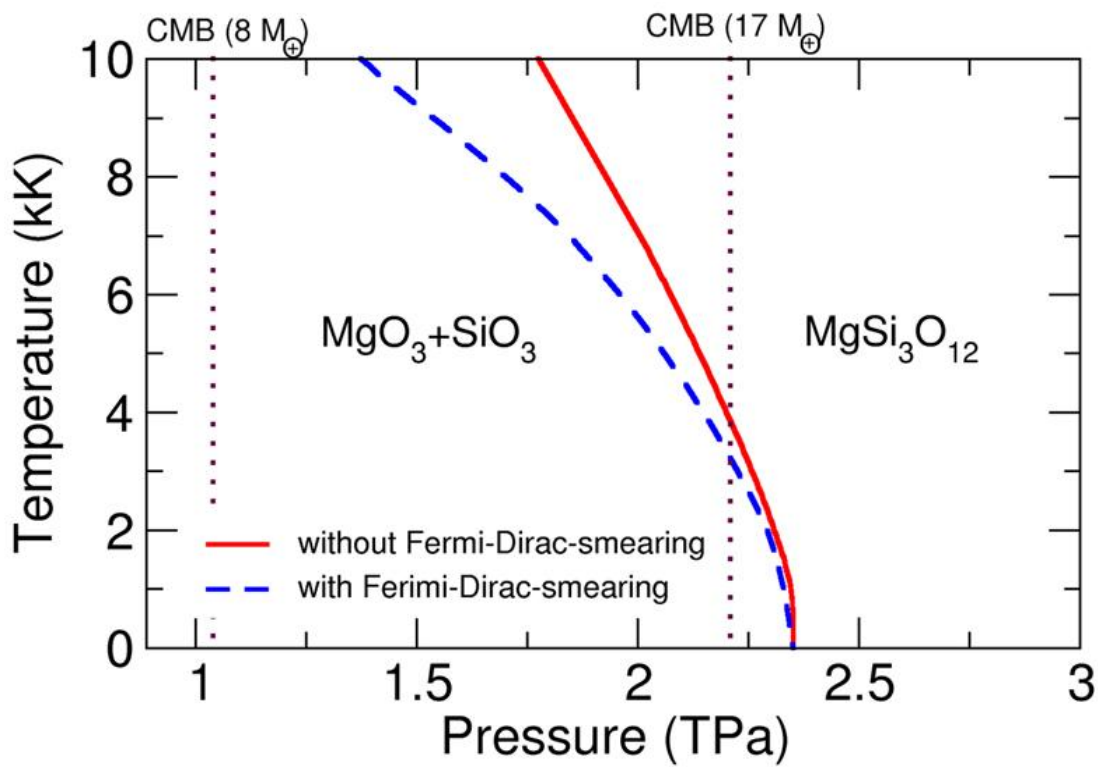

Figure 6 P-T phase diagram of MgSi$_3$O$_{12}$. The red and dotted blue lines refer to the phase boundary lines with and without Fermi-Dirac-smearing, respectively. The core-mantle boundary (CMB) pressures of super-Earths and mega-Earths with 8 and 17 M$_⊕$ are also plotted by vertical dashed lines.

## Conclusions

Using first-principles calculations and variable-composition evolutionary structure exploration in the Mg-Si-O system under exoplanet pressures, we have discovered numerous unexpected compounds. Two extraordinary compounds, $SiO_3$ and SiO, have been found to become stable at pressures above 0.51 TPa and 1.89 TPa, respectively, in the Si-O system. Both $tI$32 and $mP$16 forms of $SiO_3$ are peroxide oxides containing oxide $O^{2-}$ and peroxide $[O_2]^{2-}$ ions, while strong electron localization in the $Si_4$-tetrahedron plays the role of an additional anion to stabilize $tP$4-SiO. Besides two previously reported unusual compounds $MgO_2$ and $Mg_3O_2$, we have found another extraordinary compound, $hP$8-$MgO_3$, in the Mg-O system, which becomes thermodynamically stable at 0.89 TPa.

Taking temperature into consideration, two dissociation pathways of $MgSiO_3$ are found at relatively low (< 6.4 kK) and high (> 6,6 kK) temperature are:

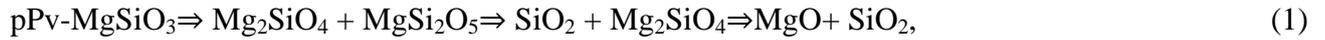

pPv-$MgSiO_3 \Rightarrow Mg_2SiO_4 + MgSi_2O_5 \Rightarrow SiO_2 + Mg_2SiO_4 \Rightarrow MgO + SiO_2$, (1)

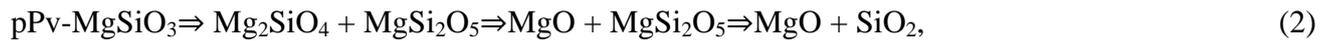

pPv-$MgSiO_3 \Rightarrow Mg_2SiO_4 + MgSi_2O_5 \Rightarrow MgO + MgSi_2O_5 \Rightarrow MgO + SiO_2$, (2)

respectively. Interestingly, besides the well-known $(MgO)_x (SiO_2)_y$ compounds, we have discovered two $(MgO_3)_x (SiO_3)_y$ compounds, $MgSi_3O_{12}$, $MgSiO_6$, which can form at 2.41 TPa and 2.95 TPa, respectively, in the Mg-Si-O system. Surprisingly, $MgSi_3O_{12}$ is predicted to be a metallic oxide with 1D-metalicity while all other oxides discussed in this work are semiconductors or insulators.

As the dissociation pathway of pPv-$MgSiO_3$ is clarified, the mineralogy and internal structure of planetary mantles can be understood much deeper. pPv-$MgSiO_3$ can survive in super-Earths with masses smaller than 6 $M_\oplus$ as shown in Fig.6b. $Mg_2SiO_4$ and $MgSi_2O_5$ can be found in the mantle of super-Earths with masses larger than 6 $M_\oplus$. Kepler-10c, 17 times heavier than Earth, would probably only have binary MgO and $SiO_2$ near the CMB. For strongly oxidized planets, $MgO_3$ and $SiO_3$ can be expected to be found. The newly discovered $MgO_3$, $SiO_3$, $MgSiO_6$, $MgSi_3O_{12}$ hold non-traditional stoichimetries, which fall off of the $MgO-SiO_2$ binary system. Given their apparent thermodynamic and lattice dynamic stability, these new compounds must be included in future models of exoplanet mineralogy in order to better understand the role that they play in massive planetary structure and evolution. The highly-oxidized $MgSi_3O_{12}$ can be formed in the lowermost of mantle of mega-Earths with masses above 20 $M_\oplus$, and a metallic layer can even exist. For O-rich planets, the extraordinary O-rich compounds $MgO_3$, $SiO_3$, $MgSi_3O_{12}$ and perhaps $MgSiO_6$ can be important planet-forming minerals. They may also appear in gas giants, as a result of reaction between Mg-silicate solid core and $H_2O$-rich fluid mantle. In future, the consideration of other important elements (e.g., Fe, Al), will likely reveal additional important high-pressure phases with similarly strange stoichiometries.

Further models of the internal structures of exoplanets must take these findings into account. Phase transitions and reactions predicted here will have a profound effect not only on the internal structure, but also on dynamical processes in planets. Exothermic reactions (with positive Clapeyron slope *dP/dT* in Fig. 4b) enhance convection, endothermic ones slow down or stop it, and a metallic layer can affect the planetary magnetic field [25, 26]. Structure, dynamics and chemistry of planetary interiors may be much more complex and surprising than previously thought.

# Computational Methods

Searches for stable compounds and structures were performed using the variable-composition evolutionary algorithm, as implemented in the USPEX code [27-31] merged with first-principles calculations within the framework of density functional theory (the Vienna Ab initio Simulation Package VASP) [32, 33] for the calculation of the total energies, relaxation of crystals, and their corresponding electronic structures. The electronic structure and force calculations at finite temperatures were implemented within the Fermi-Dirac-smearing approach [24]. The most significant feature of USPEX we used in this work is the capability of optimizing the composition and crystal structures simultaneously - as opposed to the more usual structure predictions at fixed chemical composition. The compositional search space is described via chemical building blocks. The whole range of compositions of interest is initially sampled randomly and sparsely. To ensure the child structures are within the desired area of compositional space, the chemistry-preserving constraints in the variation operators are lifted and replaced by the block correction scheme. A special "chemical transmutation" is introduced to reinforce the search efficiency. Stable compositions are determined using the convex hull construction: a compound is thermodynamically stable if the enthalpy of its decomposition into any other compounds is positive. For first-principles calculations we employed the all-electron projector augmented wave(PAW) method [34] and the generalized gradient approximation [35] for the exchange-correlation energy, along with a plane-wave cutoff energy of 800 eV and dense uniform Γ-centred k-point meshes with a reciprocal space resolution of $2\pi \times 0.03$ Å$^{-1}$. The PAW potentials have [He] core for all atoms, with radii 1.25, 1.4 and 1.15 a.u. for Mg, Si and O, respectively, which can guarantee no core overlap even at the highest pressures studied here. In addition, phonon dispersions throughout the Brillouin zone were derived using the finite-displacement approach as implemented in the Phonopy code [36]. Thermodynamic properties of these phases were investigated using their phonon spectra within the quasiharmonic approximation (QHA).

## Acknowledgements

We are grateful for support from the Government of the Russian Federation (No. 14.A12.31.0003), US National Science Foundation (EAR-1114313, DMR-1231586) and DARPA (No. W31P4Q1310005), from the "Hundred Talents Project" of Chinese Academy of Sciences, from NSFC of China (Grand Number: 51074151), as well as Beijing Supercomputing Center of CAS (including its Shenyang branch).